# Homogeneous and inhomogeneous phases in a numerical model of a time-reversal-breaking superconductor

Pedro L. Contreras E.

Address:

Departamento de Física, Universidad de Los Andes, 5101, Mérida, Venezuela

Corresponding author email: pcontreras@ula.ve

ORCID: 0000-0002-3394-1195

## ABSTRACT

In this manuscript, we find an inhomogeneous stripes phase in a numerical model for an unconventional superconductor with time-reversal-breaking symmetry and triplet odd pairing. We contrast a robust well known homogeneous phase with dilute disorder characterized by a unitary resonance and a tiny gap that resembles an s-wave superconductor, with a new inhomogeneous phase with extremely dilute disorder and real frequency stripes. This phase has only two ultra-small frequency collision values at the edges of the unitary resonance, indicating the displacement of the s-wave tiny gap to the outer edges of the resonance. We perform a statistical analysis of the real frequency density and encounter a lack of spectral stability, characterized by a sequence of vertical frequency stripes. Experimentalists can verify this finding by looking for an unstable phase in extremely clean samples of candidates for triplet superconductors.

**Keywords:** Homogeneous & Inhomogeneous Superconductor Phases. Quasi-point Nodes. Resonant Scattering Limit. Disorder, Real frequency stripes.

# 1. INTRODUCTION

The elastic scattering cross-section formalism arises from the optical theorem in non-relativistic quantum mechanics [1]. This powerful theoretical tool introduces several mechanisms serving to numerically study frequency self-consistent quantities, such as the imaginary part of the elastic scattering cross-section.

Applied to unconventional superconductors, this formalism provides insight into the physics of a self-consistent fermionic field in a complex numerical space, where real numbers represent input frequencies, and imaginary numbers represent a hidden field formed by resulting renormalized frequencies.

The physical space composed of these couple of frequencies shows the behavior of fermionic dressed quasiparticles in the unconventional superconductors, described by the average of the normal Gorkov Green Function $\langle G(k,\epsilon) \rangle$ [2] which depends on the Fourier components $k$ and $\epsilon$ [3, 4].

Henceforth, the starting point to describe the physics of unconventional superconductors in a frequency space is the self-consistent equation (1), which does not provide direct information about the Cooper pairs [5]. The bosonic information is contained in the anomalous Gorkov Green Function $F(k,\epsilon)$ [2] and serves for purposes such as the study of the non-equilibrium kinetic coefficients [3].

In this manuscript, we only study the self-consistent equation for the fermionic Green function ($GF$) that is naturally self-consistent because it is found on both sides of equation (1)

$$\langle\, G(k,\epsilon\,)\rangle = G^0(k,\epsilon) + \, G^0(k,\epsilon)\, \Sigma(k,\epsilon)\, \langle\, G(k,\epsilon)\rangle. \quad (1)$$

In equation (1), $\langle\, G(k,\epsilon\,)\rangle$ is the dressed fermionic Green function, $G^0(k,\epsilon)$ is the bare Fermionic Green function, and $\Sigma(k,\epsilon)$ is the fermionic self-energy matrix. The average $\langle\cdots\rangle_{FS}$ is over the Fermi surface.

Equation (1) is the departure to obtain a fundamental equation (2), which is suitable for a numerically rigorous self-consistent analysis including spectral characteristics. The theoretical derivation for triplet TRBS order parameter is found in [3, 6, 7] among other research literature,

$$\widetilde{\omega}(\omega + i\,0^+) = \omega + i\,\pi\,\Gamma^+ \frac{g(\widetilde{\omega})}{c^2 + g^2(\widetilde{\omega})}. \quad (2)$$

Equation (2), and the function $g(\widetilde{\omega})$ are calculated using several numerical C routines [8,9] and native codes [10].

Equation (3) is the core of the entire calculation. It is written as

$$g(\widetilde{\omega}) = g_0(\widetilde{\omega}) = \left\langle \frac{\widetilde{\omega}}{\sqrt{\widetilde{\omega}^2 - |\Delta|^2(k_x, k_y)}} \right\rangle_{FS}, \quad (3)$$

where $\widetilde{\omega}$ is a complex number, $|\Delta|^2(k_x, k_y)$ is the square of the superconducting gap in momentum space. The denominator in equation (3) shows that we subtract the bosonic part from the self-consistent frequency. Henceforth, equation (1), equation (2), and equation (3) contain only fermions information. In addition, equation (2) simplifies to a single imaginary term because for even singlet order parameter (OP) with the 1D irreducible representation $B_{1g}$ [11], and odd triplet TRSB OP with the 2D irreducible representation $E_u(\Gamma_5^-))$ [6, 7], the others terms in the Pauli matrix orthogonal space (denoted as $\boldsymbol{\tau}$) [12] become equal to zero.

The Pauli matrix orthogonal space $\boldsymbol{\tau}$ has four matrices ($\boldsymbol{\tau}_0$, $\boldsymbol{\tau}_1$, $\boldsymbol{\tau}_2$, $\boldsymbol{\tau}_3$). This space does not include magnetic effects. It is just a suitable space to represent different components of the Fermi self-energy $\Sigma(k, \epsilon)$, as Prof. G. Baym explains in his classical quantum mechanical textbook [12].

That means that the Fermi average of the function $g_1(\widetilde{\omega}) \sim \langle \Delta(k_x, k_y) \rangle_{FS} = 0$ (coupled to $\boldsymbol{\tau}_1$) takes out Cooper pairs effects for even singlet and odd triplet pairing. The average of the expression $g_3(\widetilde{\omega}) \sim \langle \xi(k_x, k_y) \rangle_{FS} = 0$ (coupled to $\boldsymbol{\tau}_3$) equals to zero, because $\xi(k_x, k_y)$ is an even function in momentum space. Thus, equation (2) has a single term coupled to $\boldsymbol{\tau}_0$, that for simplicity we denote as $g(\widetilde{\omega})$. Why is it important to analyze this term? Because it contains hidden dressed Fermi liquid effects.

We find in equation (2) two key parameters. The stoichiometric (in situ) parameter $\Gamma^+$, which quantifies disorder ($\Gamma^+ = n_{stoich}\,(\pi^2 N_F)^{-1}$). In addition, we have another parameter denoted as $c$, the inverse potential scattering strength ($c = (\pi N_F U_0)^{-1}$) [13]. These parameters are used to calculate the resonances [14], and take into account anisotropic effects [15]. Finally, the use of the irreducible representation $E_u(\Gamma_5^-)$ for the analysis of scattering effects in triplet superconductors was first due to [16], and it was applied to strontium ruthenate [17].

The imaginary term is linearly proportional to the disorder parameter $\Gamma^+$ as previously was indicated in the monograph [18]. Linearity makes $\Gamma^+$ very suitable to study numerically disorder systems, and it is able to catch up main numerical effects. The importance of the Fermi surface average is explained in the monograph [19].

Otherwise, in this type of simulations, two fundamental output physical kinetic parameters that contain self-consistent effects are the collision frequency and the inverse scattering lifetime. Their numerical values are crucial to explain the physics behind this problem. Therefore, the imaginary solution of equation (2), link these physical kinetic parameters in the following way

$$1/_\tau \, [\widetilde{\omega}(\omega + i\, 0^+)] = \nu[\widetilde{\omega}(\omega + i\, 0^+)] = 2\, \Im\, [\widetilde{\omega}(\omega + i\, 0^+)] \quad (4)$$

Equation (4) shows a direct relation among the imaginary part of the elastic scattering cross-section ($\Im$), the inverse fermionic lifetime ($\tau^{-1}$), and the collision frequency ($\nu$) of the fermionic quasiparticles.

If in equation (4), we have $\omega\, \tau(\omega) \geq 1$ for a metallic phase, this expression indicates that the singularities of the kinetic parameters become very sensitive to Fermi liquid interactions [20, 21]. But, in the resonant scattering regime for TRBS superconductors holds that $\widetilde{\omega}\, \tau\, (\widetilde{\omega}(\omega)) \geq 1$. This expression is equivalent to the relationship $\widetilde{\omega} \geq 2\, \Im\, [\widetilde{\omega}(\omega + i\, 0^+)]$.

Thus, the self-consistent analysis is sensitive to Fermi liquid interactions depending on the value of the Fermi energy $\epsilon_F$ (it tells if the OP has quasi-point or point nodes). In this context, the quasiparticles picture looks as an appropriate solution to understand equation (2).

In addition, for the fermionic quasiparticles in the resonant scattering limit, we use a relationship for the electrons free path $\ell$ and the cell parameter $a$, in other words, $\ell a^{-1} \sim 1$. This relation gave us the idea to use the TB technique, because it indicates a fermion hoping mechanism, more than a conduction mechanism.

If this analysis is used, the real axis shows what frequency window is needed for a physical interpretation due to the existence of different OP, among frequency-dependent unconventional superconductors.

For example, triplet 2D TRBS superconductors with a low transition temperature as strontium ruthenate with $T_c \sim 1.5\ K$ [17], depending strongly on in-situ disorder [22] are analyzed numerically in a real frequency window of $\pm 4$ meV [23] with quasinodal o point nodal behavior [6].

Otherwise, in HTSC superconductors with a higher transition temperature as for strontium-doped lanthanum cuprate with $T_c \sim 35\ K$ [24, 25], are analyzed in a real frequency window of $\pm 150$ meV according to recent calculations [11].

The latter is due to the problem of the frequency localization of the transition to the superconducting state from the metallic phase, and a value of the zero superconducting gap which gets incremented from its input value [11].

Henceforth, if we make the stoichiometric disorder parameter small enough in the isolate quasi-point nodal case in equation (2), we numerically obtain an inhomogeneous phase composed by vertical stripes using a 2D TRSB irreducible representation $E_u(\Gamma_5^-)$ OP, and a self-consistent calculation [26].

This work contains four sections. The introduction section aims to explain the role of fermionic dressed quasiparticles in self-consistent studies. Follows a brief section describing the appropriate parameters for the two simulations. Afterwards, the numerical results section describes and contrast the spectral density of real frequencies $\rho(\omega)$ for the two cases; and finally a concluding section summarizes this study.

**THE TIGHT BINDING MODEL**

In this manuscript, we use a well-known tight-binding (TB) model [27] to introduce two microscopic parameters in the first neighbors approximation for the average of equations (2) and (3), i.e., the Fermi energy ($\epsilon_F$) and the hopping constant ($t$), and we also need a suitable value of the zero superconducting gap ($\Delta_0 = 1.0$ meV).

The scenario where the inhomogeneous stripes phase in a time reversal breaking OP occurs with a the Fermi energy value $\epsilon_F = - 0.4$ meV that leaves a quasi-nodal gap around the $(0,\pm\pi)$ and $(\pm\pi,0)$ Brillouin points [6]. On the contrary if point nodes are considered and $\epsilon_F = - 0.04$ meV with a sharp minimum at the transition frequency [6], the inhomogeneous stripes phase does not appear.

For the two simulations, we use the rationalized Planck units $\hbar = k_B = 1$, where the real and imaginary axes are given in meV.

The model for the first neighbor TB normal state dispersion is a well-known equation

$$\xi\left(k_x, k_y\right) = \epsilon_F + 2\,t\left[\cos(k_x\,a) + \cos(k_y\,a)\right] \quad (5)$$

with the two microscopic values $\epsilon_F$ and $t$.

The irreducible representation $E_u(\Gamma_5^-)$ OP is a z-vector [6, 15, 28]

$$\mathbf{\Delta}\left(k_x, k_y\right) = \Delta_0\left[\left(\sin(k_x a) + i\,\sin(k_y a)\right]\hat{\mathbf{z}}. \quad (6)$$

The imaginary part of the equation (2) is a self-consistent frequency function in the unitary regime [3]

$$\Im\left[\widetilde{\omega}(\omega + i\,0^+)\right] = \pi\,\Gamma^+\,\frac{1}{g(\widetilde{\omega})} = \pi\,\Gamma^+\,\langle\frac{\sqrt{\widetilde{\omega}^2 - |\Delta|^2\left(k_x, k_y\right)}}{\widetilde{\omega}}\rangle_{FS}\,. \quad (7)$$

We notice in equation (7) that a divergence occurs if $\widetilde{\omega} = 0$, but for self-consistent imaginary numbers, we find that $\widetilde{\omega} > 0$ [7]. Henceforth, this simulation introduces an interesting Fermi liquid behavior subjected to a self-consistent field in an inhomogeneous stripe phase, that does not have renormalized frecuency singularities, and does not exist if the intermediate scattering regime is used.

## 3. NUMERICAL RESULTS

Figure 1 shows the first simulation using a plot with dots for the unitary case ($c = 0$), a quasi-point nodal structure where isolated nodes are located at $(0, \pm\pi)$ and $(\pm\pi, 0)$ cell points. The Fermi level with the value $\epsilon_F = -0.4$ meV (QP), a first neighbor hoping parameter $t = 0.4$ meV, a dilute in situ disorder $\Gamma^+ = 0.05$ meV, and finally a zero gap value $\Delta_0 = 1.00$ meV.

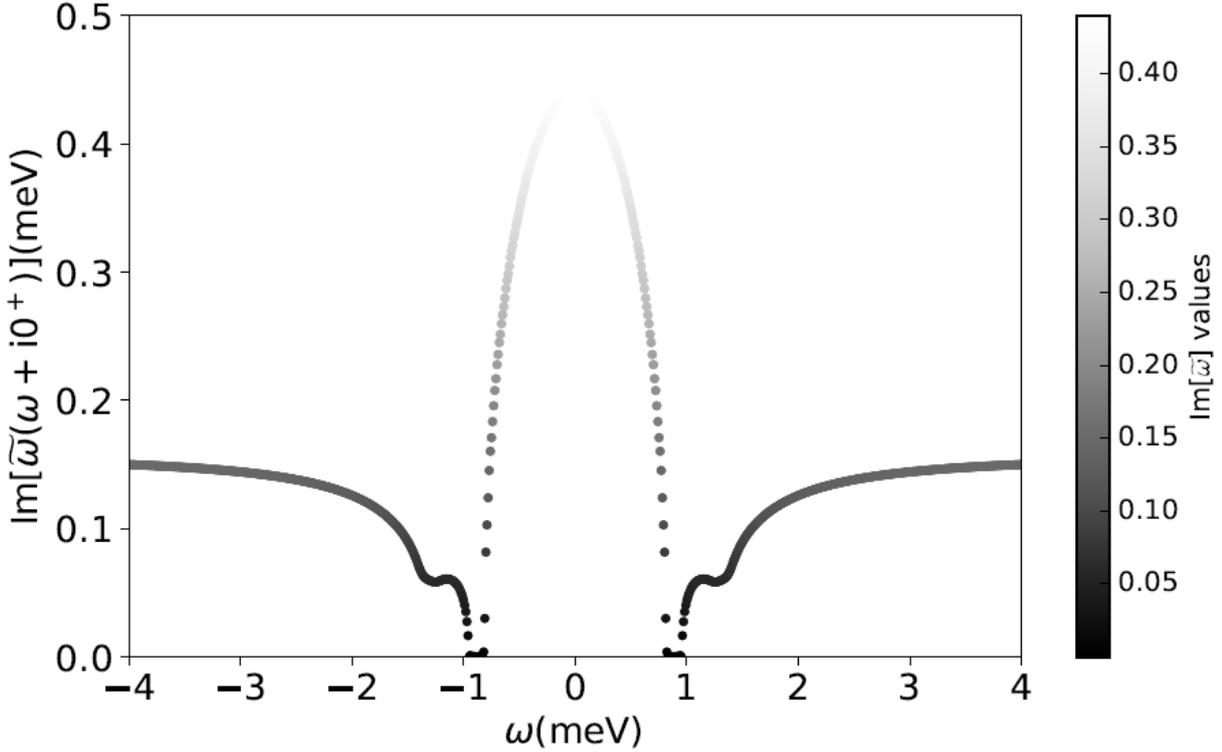

Figure 1: The scattering cross-section with dilute disorder. Please, notice how the imaginary values cover all spectrum of frequencies but the concentration of frequency points in the interval where the tiny gap locates is remarkable.

In Fig. 1, we see that from the dilute disorder case with $\Gamma^+ = 0.05$ meV, the tiny gap gives a different from zero minimum calculated with equation (7). The plot using the irred. rep. $E_u(\Gamma_5^-)$ for the OP with a **z**-vector with isolated quasi-points nodes has four different parts:

- A resonant central peak with a maximum collision frequency $\nu[\tilde{\omega}(\omega)] = 0.44$ meV.

- A tiny gap where the collision frequency varies between two orders of magnitude ($e^{-3}$ meV and $e^{-8}$ meV).

- An intermediate collision minimum $\nu[\tilde{\omega}(\omega)] = 0.06$ meV.

- Finally, a normal state metallic behavior with $\nu[\widetilde{\omega}(\omega)] = 0.15$ meV.

The tiny gap in Fig. 1 is composed by fourteen points [29], and resembles the gap of an s-wave superconductor [5]. This work will serve as a starting point to explain in a future, some controversies that recently were raised in some experiments concerning the symmetry of the OP [30, 31].

Figure 2 represents the spectral density of real points $\rho(\omega)$ according to the geometrical shape in Fig. 1. There is not a gradient in the gray color of Fig. 2, pointing out the stability of a triplet superconductor phase with a dilute disorder $\Gamma^+ = 0.05$ meV, and a zero 1.0 meV gap, concluding that the real frequency window of $\pm 4$ meV equally contributes to equation (2), nevertheless the geometrically complicate shape seen in Fig. 1.

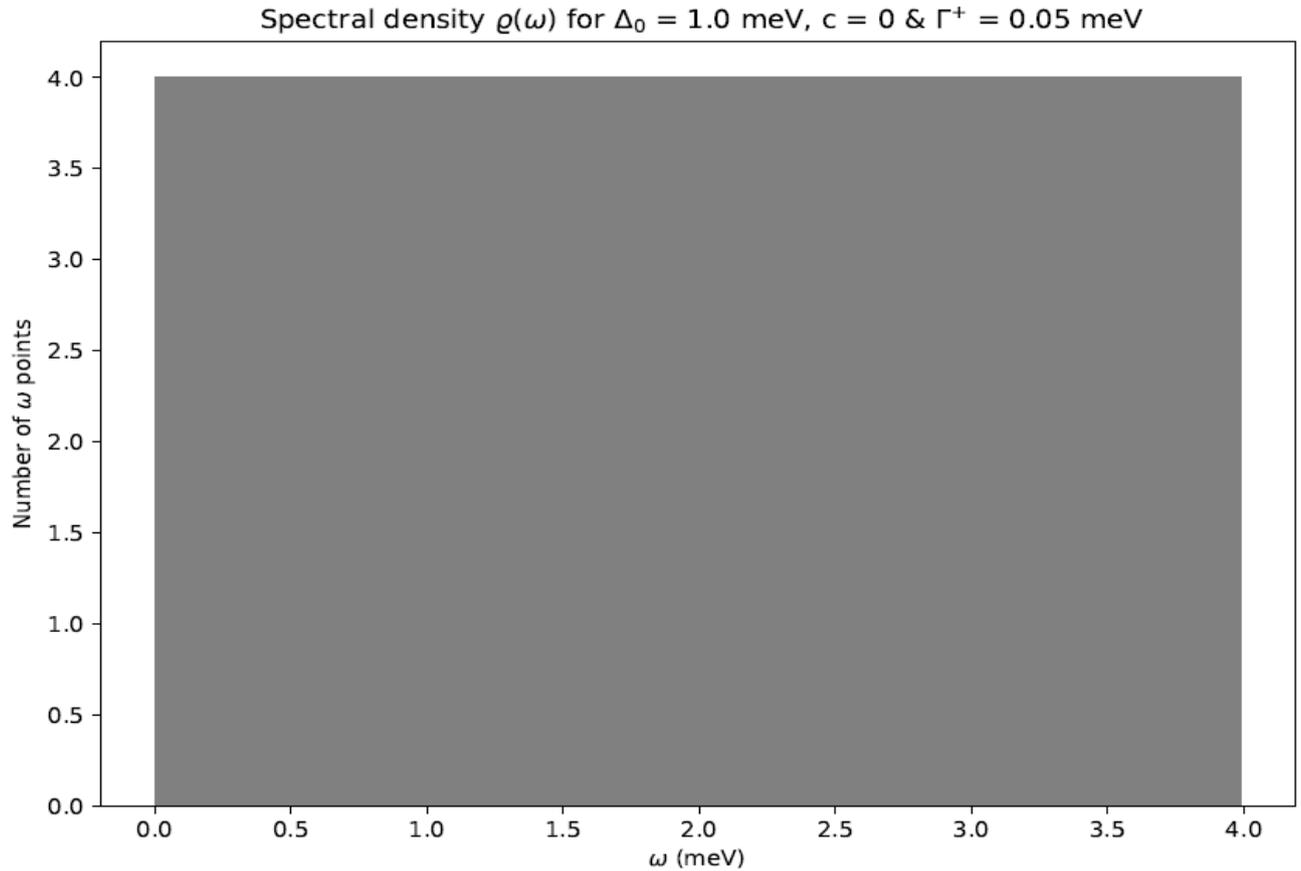

Figure 2: The spectral density $\rho(\omega)$ for the case of dilute disorder $\Gamma^+ = 0.05$ meV. The uniform gray color means that all real frequency points equally contribute to Fig. 1.

Figure 3 shows the case with a very diluted in situ disorder $\Gamma^+ = 0.01$ meV, and represents a second simulation in the unitary regimen with $\epsilon_F$ = -0.4 meV, and $t$ = 0.4 meV, the same quasi-point OP structure, and equal in value zero superconducting gap.

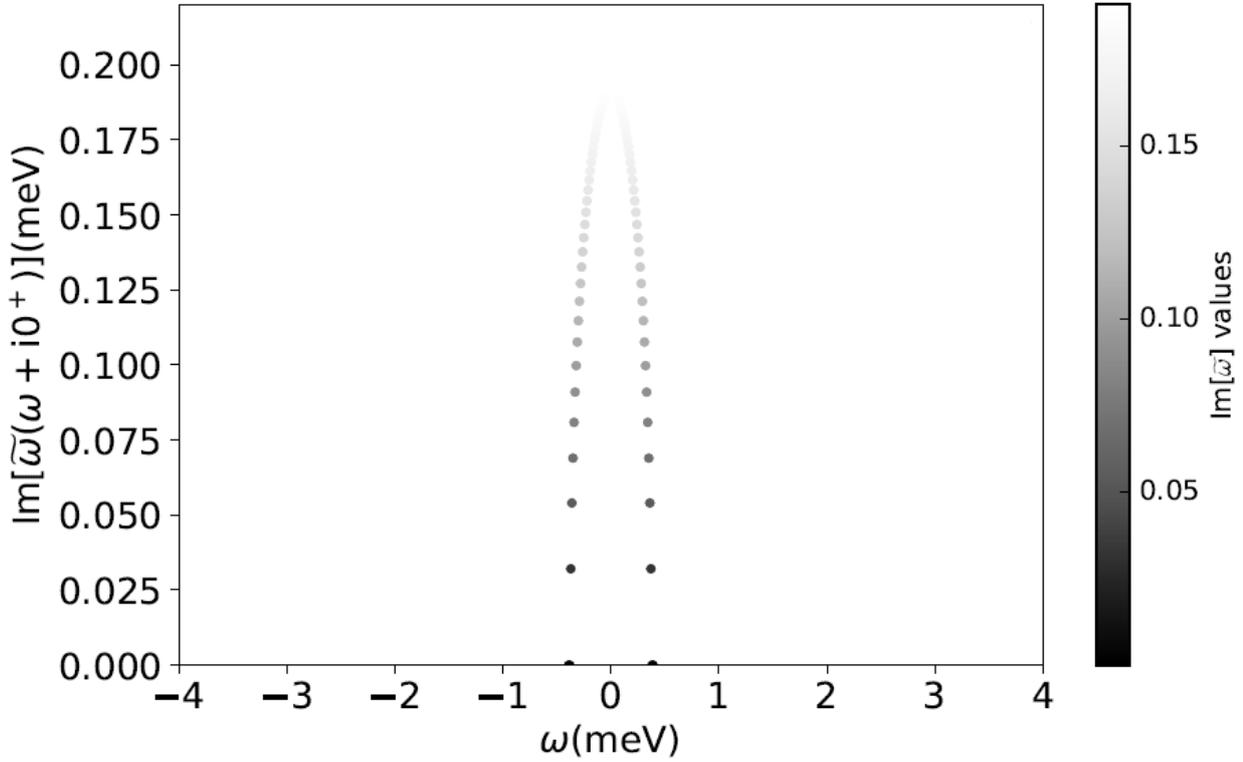

Figure 3: The scattering cross-section with very dilute disorder. Please, notice how the imaginary values cover only a small spectrum of frequencies, and the amount of frequency points on the resonance edges is seen clearly.

In Fig. 3, we notice several properties in the self-consistent simulation with very dilute disorder:

- The normal state phase vanishes, meaning that the fermionic elastic scattering cross-section does not have a normal metallic state.

- We also notice that the tiny gap is displaced to the plot edges with two extremely small points, that fall out of the range encountered in the Miyake-Narikiyo case. These collision values are $\nu[\tilde{\omega}(\omega)]$ = 4.46 x 10$^{-9}$ meV and $\nu[\tilde{\omega}(\omega)]$ = 4.06 x 10$^{-10}$ meV.

- Finally, the unitary shape in Fig. 3 has a maximum collision frequency $\nu[\tilde{\omega}(\omega)]$ = 0.38 meV.

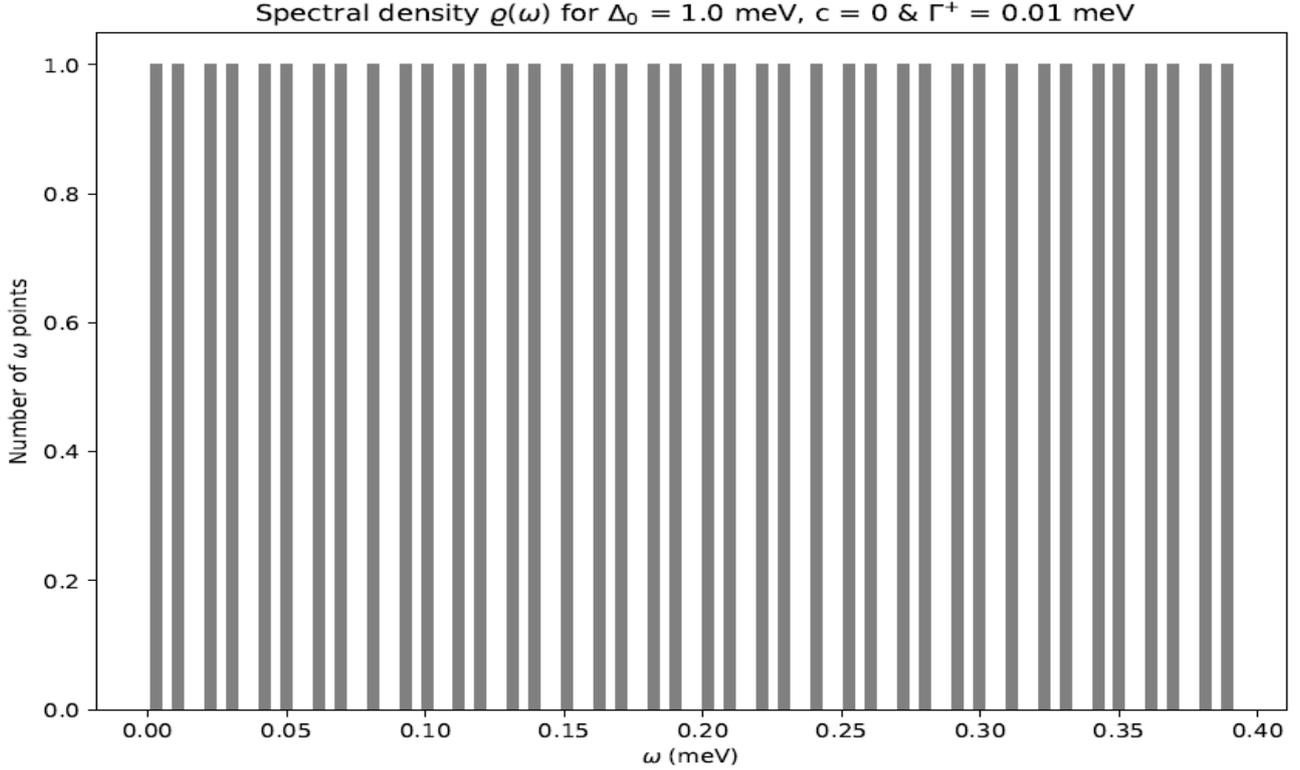

Figure 4: The inhomogeneous spectral density ρ(ω) for very dilute disorder $\Gamma^+$= 0.01 meV. The gray color represents real frequency stripes. The white empty spaces do not contribute to Fig. 3.

Figure 4 plots the spectral density $\varrho(\omega)$ for real ω with vertical stripes colored gray, coexisting with white vertical empty spaces of different width that do not contribute to the resonance.

Figure 4 points out to the existence of an anomalous nucleation of the odd triplet TRBS quasi-nodal superconductor for very dilute disorder, due to the existence of the vertical stripes.

Therefore, there is a heterogeneous spectral real density $\varrho(\omega)$, and the possibility of an insulator phase appears in Fig. 4 because the tiny gap only has two extremely low collisional frequency values, contrasting with the fourteen values in the simulation resented in Fig. 1 [29].

## 4. CONCLUSIONS

This work reports a numerical study with two phases that emerge from the use of a 2D order parameter in unconventional superconductors with the irreducible representation $E_u(\Gamma_5^-)$, and a superconducting zero gap value of 1.00 meV, only if quasi-point nodes are considered.

These phases nucleate in a simulation with the help of the elastic scattering cross-section using two values of the microscopic in situ disorder parameter. The disorder value $\Gamma^+ = 0.05$ meV for a tiny gap composed by fourteen frequency points, where we conclude that this phase is homogenous and numerically stable.

The other phase is unstable and numerically inhomogeneous due to the existence of vertical real frequency stripes, and it happens for an smaller in situ disorder value $\Gamma^+ = 0.01$ meV.

Therefore, we conclude that:

- The fermionic elastic cross-section self-consistent method is valid in the frontiers of its applicability with the emergence of a new inhomogeneous stripes phase for the irrep. $E_u(\Gamma_5^-)$, quasi-point nodal behavior, and very dilute in situ disorder.

- The resonant heterogeneous phase in real space is seen in the spectral density $\rho(\omega)$, and shows extremely low kinetic collision values.

- If real frequency stripes are present, there is not a normal metallic behavior in the scattering cross-section.

Finally, we recall that this behavior can be observed experimentally for example in extremely clean samples of superconducting strontium ruthenate because this compound is very sensitive to disorder [22].

## 5. AUTHORSHIP CONTRIBUTION STATEMENT

Pedro Contreras: Conceptualization, Methodology, Software, Investigation, Validation, Writing original draft, Supervision.

## 6. DECLARATION OF COMPETING INTEREST

The author declares that he has no known competing financial interests or personal relationships that could have appeared to influence the work reported in this paper.

# 7. REFERENCES

[1] Messiah, A. 1999. Quantum Mechanics. Dover Publications, ISBN-13-978-0-486-40924-5

[2] Gorkov L. 1958. On the energy spectrum of superconductors. Soviet Physics JETP, 7(3), 505, ISSN: 1090-6509.

[3] Mineev, V. & Samokhin, K. 1999. Introduction to Unconventional Superconductivity. Gordon and Breach Science Publishers, ISBN-90-5699-209-0

[4] Sigrist, M. & Ueda K. 1991. Phenomenological theory of unconventional superconductivity. Rev. Mod. Phys. 63(2), 239, DOI: https://doi:10.1103/revmodphys.63.239

[5] Cooper, L. 1956. Bound electron pairs in a degenerate Fermi gas. Phys. Rev. 104 (4), 1189–1190. DOI: https://doi.org/10.1103/PhysRev.104.1189

[6] Contreras, P. Osorio, D. & Tsuchiya, S. 2022. Quasi-point versus point nodes in Sr2RuO2, the case of a flat tight binding $\gamma$-sheet. Rev. Mex. Fis. 68(6):060501 DOI:

https://doi.org/10.31349/RevMexFis.68.060501

[7] Contreras, P. Osorio, D. & Devi, A. 2022. The effect of nonmagnetic disorder in the superconducting energy gap of strontium ruthenate. Physica B: Condensed Matter. 646:414330. DOI: https://doi.org/10.1016/j.physb.2022.414330

[8] Press, W., Teukolsky, S. Vetterling, W. & Flannery, B. 2002. Numerical Recipes in C: The Art of Scientific Computing, Second Edition, Cambridge University Press. ISBN-0 521 43108 5

[9] Press, W., Teukolsky, S. Vetterling, W. & Flannery, B. 2002. Example Book in C, Cambridge University Press, ISBN-0 521 43720 2

[10] Contreras, P. & Moreno, J. 2019. A non-linear minimization calculation of the renormalized frequency in dirty d-wave superconductors. Can. J. Pure Appl. Sci.. Vol. 13 (2), 4765 ISSN: 1920-3853.

[11] Contreras, P., Osorio, D. & Devi. A. 2023. Self-Consistent Study of the Superconducting Gap in the Strontium-doped Lanthanum Cuprate. Int. J. Appl. Math. Theor. Phys. Vol. 9(1), pp. 1-13. DOI: https://doi.org/10.11648/j.ijamtp.20230901.11

[12] G. Baum, Lectures on Quantum Mechanics. 1990. CRC Press, Taylor & Francis Group. ISBN-13: 978-0-8053-0667-5

[13] Schachinger, E., & Carbotte, JP. 2003. Residual absorption at zero temperature in d-wave superconductors, Phys. Rev. B 67 (2) 134509. DOI: https://doi.org/10.1103/PhysRevB.67.134509

[14] Balatsky, A., Salkola, M. & and Rosengren, A. 1995. A. Impurity induced virtual bound states in d-wave superconductors, Phys. Rev. B 51 15547.

DOI: https://doi.org/10.1103/PhysRevB.51.15547

[15] Hirschfeld, PJ. Wolfle, P. & Einzel, D. 1988. Consequences of resonant impurity scattering in anisotropic superconductors: Thermal and spin relaxation properties, Phys. Rev. B 37 83. DOI: https://doi.org/10.1103/PhysRevB.37.83

[16] Miyake, K. & Narikiyo, O. 1999. Model for Unconventional Superconductivity of Sr2RuO4: Effect of Impurity Scattering on Time-Reversal Breaking Triplet Pairing with a Tiny Gap, Phys. Rev. Lett. 83 1423. DOI: https://doi.org/10.1103/PhysRevLett.83.1423

[17] Maeno, Y., Hashimoto, H., Yoshida, K., Nishizaki, S., Fujita, T., Bednorz, J. & Lichtenberg, F. 1994. Superconductivity in a layered perovskite without copper, Nature 372 532, DOI: https://doi.org/10.1038/372532a0

[18] Lifshitz, IM., Gredeskul, S., & Pastur, L. 1988. Introduction to the theory of disordered systems. John Wiley and Sons. ISBN-10:0471875333

[19] Ziman, J. 1979. Models of Disorder: The Theoretical Physics of Homogeneously Disordered Systems, Cambridge University Press. ISBN-10-0521292808

[20] Kaganov, MI., & Contreras, P. 1994. Theory of the anomalous skin effect in metals with complicated Fermi surfaces. Soviet Physics JETP 106(6):985. ISSN: 1090-6509

[21] Walker, MB. 2001. Fermi-liquid theory for anisotropic superconductors. Phys. Rev. B. 64 (13) 134515, DOI: https://doi.org/0.1103/PhysRevB.64.134515.

[22] Mackenzie A., Haselwimmer R., Tyler A., Lonzarich G., Mori Y., Nishizaki S. & Maeno, Y. 1998. Extremely strong dependence of superconductivity on disorder in Sr2RuO4. Phys. Rev. Lett., 80:161 DOI: https://doi.org/10.1103/PhysRevLett.80.161

[23] Contreras, P. Osorio, D. & Beliayev, E. 2022. Dressed behavior of the quasiparticles lifetime in the unitary limit of two unconventional superconductors. Low Temp. Phys. 48, 187,

DOI: https://doi.org/10.1063/10.0009535

[24] Bednorz, J. & Müller. K. 1986. Possible high Tc superconductivity in the BaLaCuO system. Z. Physik B - Condensed Matter 64, 189–193. DOI: 10.1007/BF01303701

[25] Kastner, M., Birgeneau, R., Shirane, G. & Endoh, Y. 1998. Magnetic, transport, and optical properties of monolayer copper oxides Rev. Mod. Phys. 70, 897.

DOI: https://doi.org/10.1103/RevModPhys.70.897

[26] Jansen, R., Behnam Farid, and Kelly, M. 1991, The steady-state self-consistent solution to the nonlinear Wigner-function equation; a new approach, Physica B: Condensed Matter, Vol. 175(1–3):49 DOI: https://doi.org/10.1016/0921-4526(91)90688-B

[27] Harrison, WA. 1980. Electronic Structure and Properties of Solids. Dover, New York. ISBN: 9780486141787

[28] Rice TM. & Sigrist, M. 1995. Sr2RuO4: an electronic analogue of $^3$He? Journal of Physics: Condensed Matter 7 L643. DOI: http://dx.doi.org/10.1088/0953-8984/7/47/002

[29] Contreras, P. 2024. A constant self-consistent scattering lifetime in superconducting strontium ruthenate, Rev. Mex. Fís., vol. 70(6) pp. 060501 1-9 DOI:

https://doi.org/10.31349/RevMexFis.70.060501

[30] Maeno, Y., Ikeda, A. & Mattoni, G. Thirty years of puzzling superconductivity in $Sr_2RuO_4$. *Nat. Phys.* **20**, 1712–1718 (2024). DOI: https://doi.org/10.1038/s41567-024-02656-0

[31] Leggett, A. & Liu, Y. 2021. Symmetry Properties of Superconducting Order Parameter in Sr2RuO4. J Supercond. Nov. Magn. 34:1647. https://doi.org/10.1007/s10948-020-05717-6